\documentclass[preprint,prd,showpacs,superscriptaddress,preprintnumbers,floatfix]{revtex4}

\usepackage{amsmath}
\usepackage{amssymb}
\usepackage{enumerate}
\usepackage{bm}
\usepackage{dcolumn}
\usepackage{graphicx}
\newcommand{\gtwo}{$g\!-\!2$}
\newcommand{\LB}{L\!B}

\newcommand{\ph}{\phantom{+}}

\begin{document}


\preprint{RIKEN-TH-192}

\title{%
Tenth-order lepton $g\!-\!2$: Contribution of some fourth-order
radiative corrections to the sixth-order $g\!-\!2$ containing
 light-by-light-scattering subdiagrams 
}


\author{T.~Aoyama}
\affiliation{Kobayashi-Maskawa Institute for the Origin of Particles and the Universe (KMI), Nagoya University, Nagoya, 464-8602, Japan }
\affiliation{Theoretical Physics Laboratory, Nishina Center, RIKEN, Wako, 351-0198, Japan }

\author{M.~Hayakawa}
\affiliation{Kobayashi-Maskawa Institute for the Origin of Particles and the Universe (KMI), Nagoya University, Nagoya, 464-8602, Japan }
\affiliation{Theoretical Physics Laboratory, Nishina Center, RIKEN, Wako, 351-0198, Japan }
\affiliation{Department of Physics, Nagoya University, Nagoya, 464-8602, Japan }

\author{T.~Kinoshita}
\affiliation{Theoretical Physics Laboratory, Nishina Center, RIKEN, Wako, 351-0198, Japan }
\affiliation{Laboratory for Elementary Particle Physics, Cornell University, Ithaca, New York, 14853, U.S.A }

\author{M.~Nio}
\affiliation{Theoretical Physics Laboratory, Nishina Center, RIKEN, Wako, 351-0198, Japan }

\begin{abstract}
This paper reports the tenth-order QED contribution to lepton \gtwo\ 
from diagrams of three gauge-invariant sets VI(d), VI(g), and VI(h), which 
are obtained by including various fourth-order radiative corrections
to the sixth-order \gtwo\ containing light-by-light-scattering subdiagrams. 
In the case of electron \gtwo, they consist of 492, 480, and 630 
vertex Feynman diagrams, respectively. 
The results of numerical integration, including mass-dependent terms 
containing muon loops, are 
$ 1.8418~(95)~(\alpha/\pi)^5$ for the Set VI(d), 
$-1.5918~(65)~(\alpha/\pi)^5$ for the Set VI(g), 
and
$ 0.1797~(40)~(\alpha/\pi)^5$ for the Set VI(h), 
respectively. 
We also report the contributions to the muon \gtwo, which derive 
from diagrams containing an electron, muon or tau lepton loop: 
Their sums are 
$ -5.876~(802)~(\alpha/\pi)^5$ for the Set VI(d), 
$  5.710~(490)~(\alpha/\pi)^5$ for the Set VI(g), and
$ -8.361~(232)~(\alpha/\pi)^5$ for the Set VI(h), respectively. 
\end{abstract}

%

\maketitle

\section{Introduction}
\label{sec:intro}

The anomalous magnetic moment \gtwo\ of the electron is one of the 
most vigorously studied physical quantities at present, which 
provides a very stringent test of the validity of 
quantum electrodynamics (QED). 
To match the precision of the latest measurement of electron \gtwo
\cite{Hanneke:2008tm} 
the theory must include the QED radiative correction up to 
the eighth order 
\cite{Kinoshita:2005zr,Aoyama:2007dv,Aoyama:2007mn} 
as well as the hadronic contribution 
\cite{Hagiwara:2006jt,Jegerlehner:2009ry,Davier:2009zi,Krause:1996rf,Melnikov:2003xd,Bijnens:2007pz,Prades:2009tw,Nyffeler:2009tw} 
and the electroweak contribution 
\cite{Czarnecki:1996ww,Knecht:2002hr,Czarnecki:2002nt} 
within the context of the standard model. 
As a matter of fact, the largest theoretical uncertainty now comes from 
the tenth-order QED contribution which has not yet been evaluated
and is given only a crude estimate \cite{Mohr:2008fa}.
Thus it is an urgent matter to evaluate 
the actual value of the tenth-order term. 
To accomplish this task we started a systematic program 
several years ago to evaluate the complete tenth-order contribution
\cite{Kinoshita:2005sm,Aoyama:2005kf,Aoyama:2007bs,Aoyama:2008gy,Aoyama:2008hz,Aoyama:2010yt}. 

The tenth-order QED contribution to the electron \gtwo\ consists of 
the mass-independent term $A_1^{(10)}$ and 
the mass-dependent terms $A_2^{(10)}$ and $A_3^{(10)}$ 
in which muon and/or tau lepton loop is involved, which  may be expressed as 
\begin{equation}
  a_e^{(10)} = \left[
    A_1^{(10)}
    + A_2^{(10)}(m_e/m_\mu) + A_2^{(10)}(m_e/m_\tau)
    + A_3^{(10)}(m_e/m_\mu,m_e/m_\tau)
    \right] \left(\frac{\alpha}{\pi}\right)^5.
\end{equation}
The mass-independent term $A_1^{(10)}$ may be classified 
into six sets and further divided into 32 gauge-invariant subsets 
according to the type of the closed lepton loop subdiagram. 
Thus far, the numerical evaluation of 21 subsets has been carried out 
and the results were published 
\cite{Kinoshita:2005sm,Aoyama:2005kf,Aoyama:2007bs,Aoyama:2008gy,Aoyama:2008hz,Aoyama:2010yt}. 

\begin{figure}
  \begin{center}
    \includegraphics[scale=1.2]{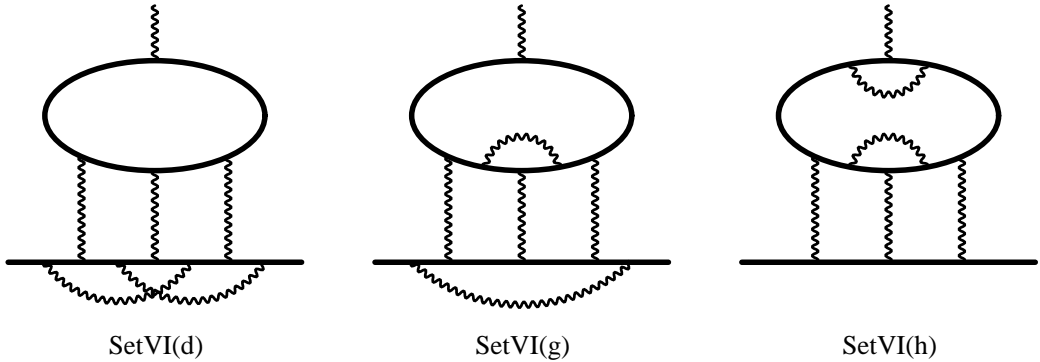}
  \end{center}
  \caption{%
    \label{fig:set6dgh}
    Typical diagrams of Set VI(d), Set VI(g), and Set VI(h). 
}
\end{figure}

In this paper we focus our attention on the 
gauge-invariant set VI which consists of all diagrams 
containing a light-by-light-scattering subdiagram,
one of whose photon vertex is external. 
(We call this an \textit{external} light-by-light-scattering subdiagram.) 
Of eleven gauge-invariant subsets of the Set VI,
eight have been evaluated previously \cite{Kinoshita:2005sm}.
The purpose of this paper is to report the evaluation of
the remaining three gauge-invariant subsets: 
Sets VI(d), VI(g), and VI(h). 
In diagrams of Set VI(d) two virtual photon lines are attached to the open 
lepton line. This set contains 492 vertex diagrams. 
In diagrams of Set VI(g) one virtual photon line is attached to the open 
lepton line and the other virtual photon line is attached to the closed lepton 
loop. This set contains 480 vertex diagrams. 
In diagrams of Set VI(h) two virtual photon lines are attached to the closed 
lepton loop. This set contains 630 vertex diagrams. 
Typical diagrams of these sets are shown in Fig.~\ref{fig:set6dgh}. 

Our numerical evaluation of Feynman diagrams is based on 
the parametric integration formula 
\cite{Cvitanovic:1974uf,Cvitanovic:1974sv,Kinoshita:1990}. 
To handle a relatively large number of diagrams systematically without 
errors, we developed an automated code-generating system 
called \textsc{gencode\textrm{LL}N}
that produces FORTRAN codes for the numerical integration. 
It is an adaptation of the previously developed system for the type of 
diagrams without lepton loops 
\cite{Aoyama:2005kf,Aoyama:2007bs} 
to the diagrams containing an external light-by-light-scattering subdiagram. 

This paper is organized as follows. 
Section~\ref{sec:scheme} describes our scheme for numerical evaluation. 
Section~\ref{sec:result} gives the results of the numerical evaluation. 
Section~\ref{sec:discussion} is devoted to the summary and discussion. 
In Appendix~\ref{sec:loopmatrix} we describes an algorithm for identifying 
independent set of loops on a diagram that is required for constructing 
the amplitude. 
For simplicity the factor $(\alpha/\pi)^5$ is omitted 
in Secs.~\ref{sec:scheme} and~\ref{sec:result}. 

\section{Numerical evaluation scheme}
\label{sec:scheme}

In this section we describe our scheme for the numerical evaluation 
of the diagrams of Sets VI(d), VI(g), and VI(h). 
The diagram that belongs to these sets consists of 
an open lepton line ($\ell_1$) and a closed lepton loop ($\ell_2$) 
that forms a light-by-light-scattering (\textit{l-by-l}) subdiagram, 
where $\ell_1$ and $\ell_2$ refer to the types of leptons, 
i.e. electron ($e$), muon ($m$), or tau lepton ($t$). 
The mass-dependence of these diagrams and amplitudes is characterized by
($\ell_1$,$\ell_2$) or by superscript ${}^{(\ell_1\ell_2)}$.

\begin{figure*}
  \begin{center}
    \includegraphics[scale=0.75]{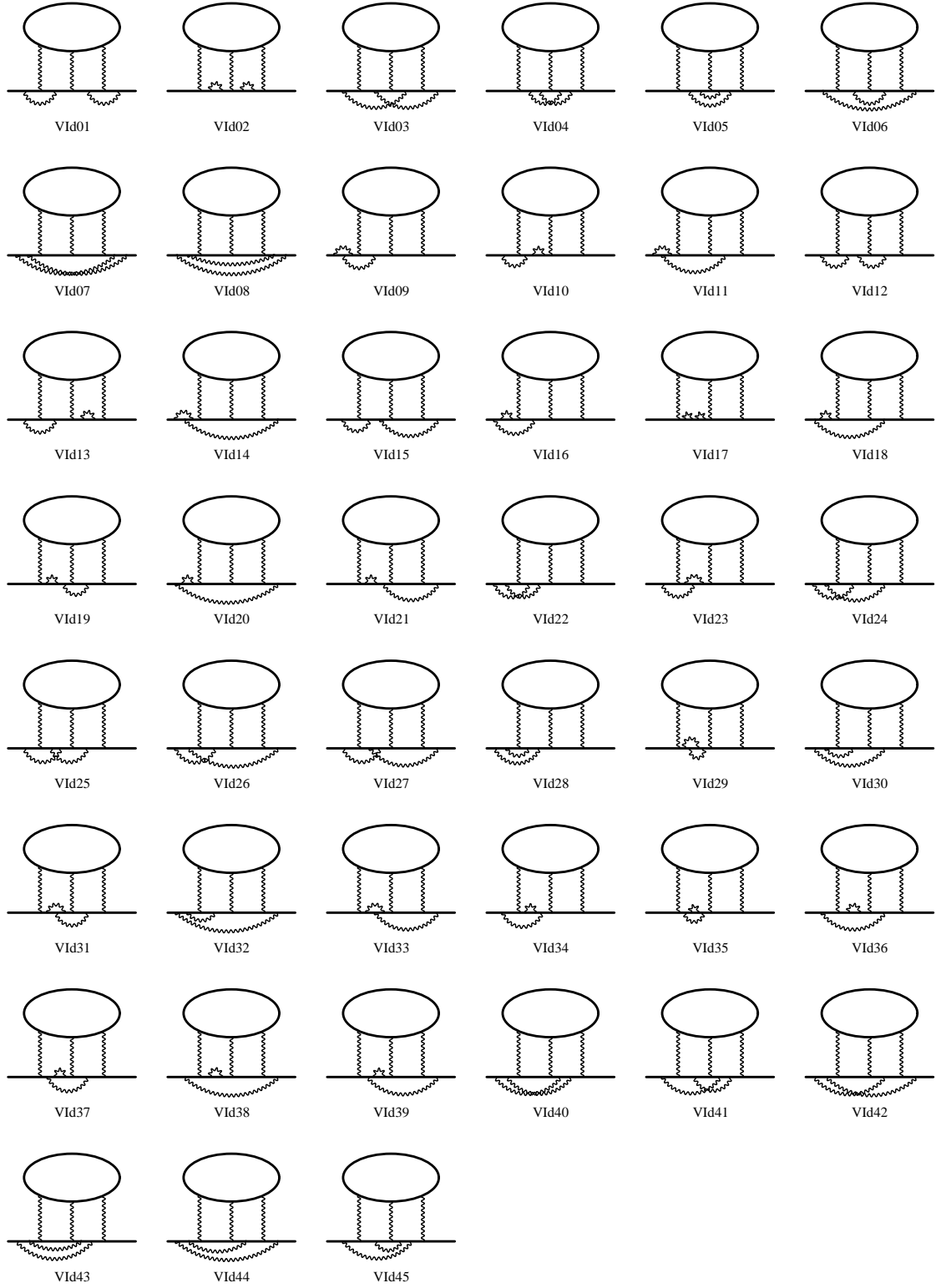}
  \end{center}
  \caption{%
    \label{fig:set6d}
    The contribution of Set VI(d) 
    is represented by 45 independent diagrams as listed. 
}
\end{figure*}

\begin{figure*}
  \begin{center}
    \includegraphics[scale=0.75]{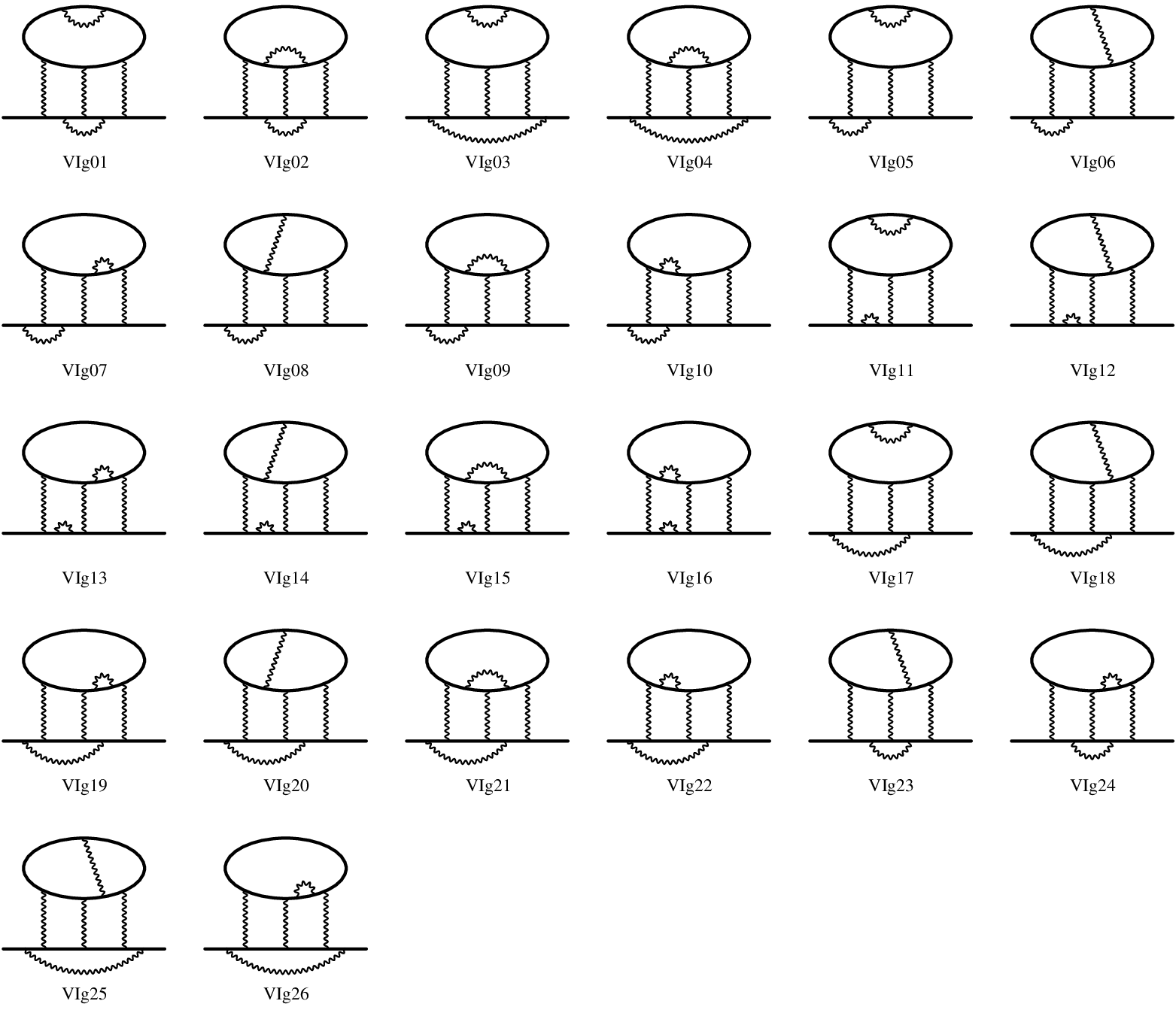}
  \end{center}
  \caption{%
    \label{fig:set6g}
    The contribution of Set VI(g) 
    is represented by 26 independent diagrams as listed. 
}
\end{figure*}

\begin{figure*}
  \begin{center}
    \includegraphics[scale=0.75]{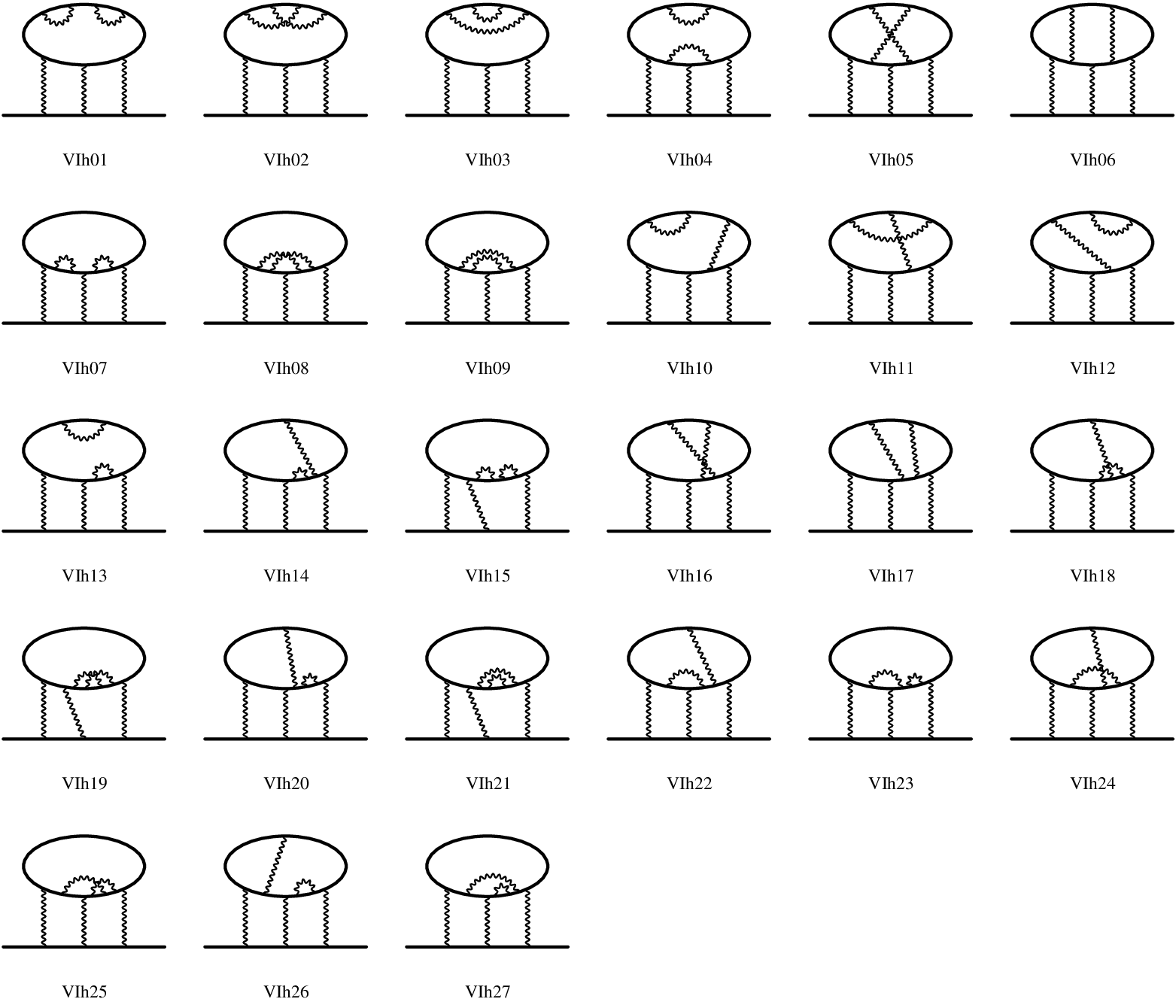}
  \end{center}
  \caption{%
    \label{fig:set6h}
    The contribution of Set VI(h) 
    is represented by 27 independent diagrams as listed. 
}
\end{figure*}

We adopt a relation derived from the Ward-Takahashi identity 
\begin{equation}
  \Lambda^\nu(p,q) \simeq 
  - q^\mu \left.\frac{\partial\Lambda_\mu(p,q)}{\partial q_\nu}\right|_{q\to 0}
  - \frac{\partial\Sigma(p)}{\partial p_\nu}
  \label{eq:wt}
\end{equation}
where $\Lambda^\nu(p,q)$ is the sum of proper vertex parts which are obtained
by inserting an external photon vertex in the lepton lines of 
the self-energy function $\Sigma(p)$ of a diagram $\mathcal{G}$ 
in all possible ways.
Taking account of the charge conjugation and time reversal symmetry 
the numbers of independent integrals to evaluate become 
45 for Set VI(d) (see Fig.~\ref{fig:set6d}), 
26 for Set VI(g) (see Fig.~\ref{fig:set6g}), and 
27 for Set VI(h) (see Fig.~\ref{fig:set6h}). 
%

The amplitude of a Feynman diagram is turned into an integral 
over the Feynman parameters assigned to the lepton and photon lines 
by using the parametric integral formula 
\cite{Cvitanovic:1974uf}. 
Note that the contribution of the second term on the right-hand side 
of Eq.~(\ref{eq:wt}) vanishes due to the Furry's theorem. 

In our numerical procedure the renormalization of the amplitude 
is carried out by the subtractive renormalization. 
The unrenormalized amplitude $M_\mathcal{G}$ of a diagram $\mathcal{G}$ 
is related to a finite calculable quantity $\Delta M_\mathcal{G}$ 
by appropriate subtraction terms of UV and IR divergences. 
These subtraction terms are prepared in the form of integrals 
over the same Feynman parameter space so that 
they cancel out the divergent behavior of 
the original unrenormalized integral {\it point-by-point}. 

The UV divergence arising from the l-by-l subdiagram must be 
regularized, e.g., by the Pauli-Villars regularization. 
However, in Eq.~(\ref{eq:wt}) the Ward-Takahashi-summed amplitude 
is given as the differentiation of $\Lambda^\mu(p,q)$ 
with respect to $q_\nu$. 
Therefore the divergence from the l-by-l loop is lifted, 
and the PV regularization is no longer needed. 

As a consequence the source of the UV divergences 
resides only in the vertex and 
self-energy subdiagrams of second and fourth order.
These divergences are handled by {\em K}~operation 
\cite{Cvitanovic:1974sv,Kinoshita:1981ww}. 
By definition the {\em K}~operation yields the subtraction integral 
which is analytically factorizable into a product or a sum of products 
of lower-order quantities. 
It is symbolically denoted by 
\begin{equation}
  \mathbb{K}_{\mathcal{S}} M_{\mathcal{G}} 
  = 
  L_{\mathcal{S}}^{\text{UV}} M_{\mathcal{G}/\mathcal{S}}, 
\end{equation}
when $S$ is a vertex subdiagram, and by 
\begin{equation}
  \mathbb{K}_{\mathcal{S}} M_{\mathcal{G}} 
  = 
  {\delta m}_{\mathcal{S}}^{\text{UV}} M_{\mathcal{G}/\mathcal{S}} 
  +
  B_{\mathcal{S}}^{\text{UV}} M_{[\mathcal{G}/\mathcal{S},i]}, 
\end{equation}
when $S$ is a self-energy subdiagram. 
Here, the superscript $\text{UV}$ means that 
the leading UV divergent part is taken for 
the vertex renormalization constant $L$, 
the mass renormalization constant $\delta m$, and 
the wave-function renormalization constant $B$, respectively. 
Note that ${\delta m}_2^{\text{UV}} = {\delta m}_2$. 
We also apply {\em R}~subtraction 
\cite{Aoyama:2007bs}
by which the residual part 
$\widetilde{\delta m} \equiv {\delta m} - {\delta m}^{\text{UV}}$ 
of the fourth-order mass renormalization constant is subtracted 
away to accomplish complete subtraction of $\delta m$. 

Some diagrams of Set VI(d) and Set VI(g) have IR divergences. 
For example, the diagram VIg04 of Set VI(g) shown in 
Fig.~\ref{fig:set6g}, 
in which a photon line attached to the open lepton line at both ends 
encloses an eighth-order l-by-l subdiagram, 
exhibits an IR divergence when this outermost photon goes soft. 
These divergences are subtracted away by {\em I}~subtraction 
\cite{Aoyama:2007bs}. 
By construction the {\em I} subtraction term factorizes as
\begin{equation}
  \mathbb{I}_{\mathcal{S}} M_{\mathcal{G}} 
  = 
  M_{\mathcal{S}} \widetilde{L}_{\mathcal{G}/\mathcal{S},i}, 
\end{equation}
where $\widetilde{L} \equiv L - L^{\text{UV}}$ denotes the 
residual part of the vertex renormalization constant. 

The finite amplitude $\Delta M_{\mathcal{G}}$ obtained so far 
differs from the standard renormalized quantity, 
because the subtraction terms involve only a fraction of 
the renormalization constants relevant to the divergences. 
To achieve the standard on-the-mass-shell renormalization, 
the differences are collected over the diagrams of the subset, 
which is finite,
and added to $\Delta M_{\mathcal{G}}$.
This step is called the residual renormalization. 

The exactly renormalized contributions of Sets VI(d), VI(g), and VI(h) 
to the magnetic moment are given by the formulas: 
\begin{align}
  \label{eq:res-set6d}
  a_{\ell_1}^{(10)}[\text{VI(d)}^{(\ell_1\ell_2)}] 
  & =
  {\Delta M}_{\text{VI(d)}}^{(\ell_1\ell_2)}
  - 4 \Delta\LB_{2}\,{\Delta M}_{\text{IVc}}^{(\ell_1\ell_2)}
  + \left( - 2 \Delta\LB_{4} + 5 (\Delta\LB_{2})^2 \right)\,{a}_{\text{6LL}}^{(\ell_1\ell_2)} \nonumber \\[1ex]
  & =
  \Delta M_{\text{VI(d)}}^{(\ell_1\ell_2)}
  - 4 \Delta\LB_{2}\,a_{\text{IVc}}^{(\ell_1\ell_2)}
  - \left( 2 \Delta\LB_{4} + 3 (\Delta\LB_{2})^2 \right)\,a_{\text{6LL}}^{(\ell_1\ell_2)}, \\[1ex]
  \label{eq:res-set6g}
  a_{\ell_1}^{(10)}[\text{VI(g)}^{(\ell_1\ell_2)}] 
  & =
  \Delta M_{\text{VI(g)}}^{(\ell_1\ell_2)}
  - 2 \Delta\LB_{2}\,{\Delta M}_{\text{IVb}}^{(\ell_1\ell_2)}
  - 3 \Delta\LB_{2}\,{\Delta M}_{\text{IVc}}^{(\ell_1\ell_2)}
  + 6 (\Delta\LB_{2})^2\,{a}_{\text{6LL}}^{(\ell_1\ell_2)} \nonumber \\[1ex]
  & =
  \Delta M_{\text{VI(g)}}^{(\ell_1\ell_2)}
  - 2 \Delta\LB_{2}\,a_{\text{IVb}}^{(\ell_1\ell_2)}
  - 3 \Delta\LB_{2}\,a_{\text{IVc}}^{(\ell_1\ell_2)}
  - 6 (\Delta\LB_{2})^2\,a_{\text{6LL}}^{(\ell_1\ell_2)}, \\[1ex]
  \label{eq:res-set6h}
  a_{\ell_1}^{(10)}[\text{VI(h)}^{(\ell_1\ell_2)}] 
  & =
  \Delta M_{\text{VI(h)}}^{(\ell_1\ell_2)}
  - 5 \Delta\LB_{2}\,{\Delta M}_{\text{IVb}}^{(\ell_1\ell_2)}
  + \left( - 3 \Delta\LB_{4} + 9 (\Delta\LB_{2})^2 \right)\,{a}_{\text{6LL}}^{(\ell_1\ell_2)} \nonumber \\[1ex]
  & =
  \Delta M_{\text{VI(h)}}^{(\ell_1\ell_2)}
  - 5 \Delta\LB_{2}\,a_{\text{IVb}}^{(\ell_1\ell_2)}
  - \left( 3 \Delta\LB_{4} + 6 (\Delta\LB_{2})^2 \right)\,a_{\text{6LL}}^{(\ell_1\ell_2)}. 
\end{align}
Here, 
${\Delta M}_{\text{VI(d)}}$, 
${\Delta M}_{\text{VI(g)}}$, and 
${\Delta M}_{\text{VI(h)}}$ are the sum of the finite amplitude of the 
diagrams within the subsets VI(d), VI(g), and VI(h), respectively. 
$a_{\text{6LL}}$ is the sixth-order anomalous magnetic moment 
containing the fourth-order l-by-l diagram. 
$a_{\text{IVb}}$ and $a_{\text{IVc}}$ are the eighth-order anomalous 
magnetic moments of the set of diagrams containing 
the external l-by-l subdiagram with a virtual photon line attached 
to the lepton loop (IVb) or to the open lepton line (IVc). 
${\Delta M}_{\text{IVb}}$ and ${\Delta M}_{\text{IVc}}$ 
are their finite part defined in Ref.~\cite{Kinoshita:1981ww}. 
$\Delta\LB_2$ and $\Delta\LB_4$ are the sum of the finite part of 
vertex and wave-function renormalization constants 
of second and fourth order, respectively. 

The code-generating program \textsc{gencode\textrm{LL}N} 
takes a one-line representation of a diagram as an input, and generates
the numerical integration program formatted in FORTRAN. 
During this process it finds the form of the unrenormalized amplitudes, 
identifies the divergence structure, and constructs the UV- and/or IR-subtraction integrals. 
The symbolic manipulations concerning e.g the gamma matrix calculus 
and the analytic integration using homemade integration tables 
are processed with the helps of FORM \cite{Vermaseren:2000nd} and Maple. 

\section{Results}
\label{sec:result}

The numerical integration is carried out using the adaptive-iterative
Monte-Carlo integration routine VEGAS \cite{vegas}. 
The numerical values of individual amplitudes of 
Set VI(d), Set VI(g), and Set VI(h) are listed in 
Tables~\ref{table:set6d},~\ref{table:set6g}, and~\ref{table:set6h}, 
respectively, 
for the mass-independent term and the mass-dependent terms in which 
$(\ell_1,\ell_2) = (e,m)$, $(m,e)$, and $(m,t)$. 
We use the muon-electron mass ratio $m_\mu/m_e = 206.768~282~3~(52)$
and the tau-muon mass ratio $m_\tau/m_\mu = 16.818~3~(27)$ for 
numerical evaluation \cite{Mohr:2008fa}.

\subsection{Mass-independent contribution}
\label{sec:result:a1}

Let us first consider the case in which $\ell_1$ and $\ell_2$ are 
of the same type of lepton, i.e., 
$\ell_1 = \ell_2 = e, m, \text{or}\ t$. 
This gives a mass-independent contribution to the lepton \gtwo. 
The numerical values are listed in the second columns of 
Table~\ref{table:set6d} for Set VI(d), 
Table~\ref{table:set6g} for Set VI(g), and
Table~\ref{table:set6h} for Set VI(h), respectively. 
The values of the sixth- and eighth-order amplitudes and the 
finite renormalization constants are listed in 
Table~\ref{table:residual}. 
Putting these values into 
Eqs.~(\ref{eq:res-set6d}),~(\ref{eq:res-set6g}), and~(\ref{eq:res-set6h}), 
the mass-independent contributions $A_1$ of the respective subsets are: 
\begin{align}
  \label{eq:a1-set6d}
  A_1^{(10)} [\text{Set VI(d)}] &= \ph 1.840~5~(95), \\[1ex]
  \label{eq:a1-set6g}
  A_1^{(10)} [\text{Set VI(g)}] &=    -1.591~3~(65),  \\[1ex] 
  \label{eq:a1-set6h}
  A_1^{(10)} [\text{Set VI(h)}] &= \ph 0.179~7~(40). 
\end{align}

\subsection{Mass-dependent contribution ({\it e,m})}
\label{sec:result:a2-em}

The mass-dependent contribution to the electron \gtwo\ in which 
the light-by-light-scattering subdiagram consists of the muon loop, 
i.e., $\ell_1 = e$ and $\ell_2 = m$, is found 
from the numerical values listed in the third columns of 
Table~\ref{table:set6d} for Set VI(d), 
Table~\ref{table:set6g} for Set VI(g), and
Table~\ref{table:set6h} for Set VI(h), respectively. 
The values of the mass-dependent sixth- and eighth-order amplitudes are 
listed in Table~\ref{table:residual}. 
Putting these values into 
Eqs.~(\ref{eq:res-set6d}),~(\ref{eq:res-set6g}), and~(\ref{eq:res-set6h}), 
we obtain the mass-dependent contributions $A_2(m_e/m_\mu)$ 
of the respective subsets: 
\begin{align}
  \label{eq:a2-em-set6d}
  A_2^{(10)}(m_e/m_\mu) [\text{Set VI(d)}] &= \ph 0.001~276~(76), \\[1ex]
  \label{eq:a2-em-set6g}
  A_2^{(10)}(m_e/m_\mu) [\text{Set VI(g)}] &=    -0.000~497~(29), \\[1ex]
  \label{eq:a2-em-set6h}
  A_2^{(10)}(m_e/m_\mu) [\text{Set VI(h)}] &= \ph 0.000~045~(10). 
\end{align}

\subsection{Mass-dependent contribution ({\it m,e})}
\label{sec:result:a2-me}

Similarly, the mass-dependent contribution to the muon \gtwo\ in which 
the light-by-light-scattering subdiagram consists of the electron loop, 
i.e., $\ell_1 = m$ and $\ell_2 = e$, is found 
from the numerical values listed in the fourth column of 
Tables~\ref{table:set6d},~\ref{table:set6g}, and~\ref{table:set6h}. 
Their contributions are: 
\begin{align}
  \label{eq:a2-me-set6d}
  A_2^{(10)}(m_\mu/m_e) [\text{Set VI(d)}] &=    -7.798~(801), \\[1ex]
  \label{eq:a2-me-set6g}
  A_2^{(10)}(m_\mu/m_e) [\text{Set VI(g)}] &= \ph 7.346~(489), \\[1ex]
  \label{eq:a2-me-set6h}
  A_2^{(10)}(m_\mu/m_e) [\text{Set VI(h)}] &=    -8.546~(231). 
\end{align}

\subsection{Mass-dependent contribution ({\it m,t})}
\label{sec:result:a2-mt}

The mass-dependent contribution of the tau-lepton loop 
to the muon \gtwo\ is also evaluated and their numerical values are 
listed in the fifth column of 
Tables~\ref{table:set6d},~\ref{table:set6g}, and~\ref{table:set6h}. 
The results are 
\begin{align}
  \label{eq:a2-mt-set6d}
  A_2^{(10)}(m_\mu/m_\tau) [\text{Set VI(d)}] &= \ph 0.081~77~(161), \\[1ex]
  \label{eq:a2-mt-set6g}
  A_2^{(10)}(m_\mu/m_\tau) [\text{Set VI(g)}] &=    -0.044~51~( 96), \\[1ex]
  \label{eq:a2-mt-set6h}
  A_2^{(10)}(m_\mu/m_\tau) [\text{Set VI(h)}] &= \ph 0.004~85~( 46). 
\end{align}

\begingroup
\squeezetable
\renewcommand{\baselinestretch}{1.2}

\begin{table*}
  \caption{%
Numerical values of diagrams of Set VI(d) 
for the mass-independent contributions $(e,e)$ 
and the mass-dependent contributions $(e,m)$, $(m,e)$, and $(m,t)$. 
The numerals in the parentheses show the uncertainty of the numerical 
integration carried out using VEGAS. 
The number of sampling points is $10^9$ times approximately 100 iterations. 
  \label{table:set6d}
}
  \begin{ruledtabular}
    \begin{tabular}{ldddd}
\multicolumn{1}{c}{Diagram} & 
\multicolumn{1}{c}{$\Delta M^{(ee)}$} &
\multicolumn{1}{c}{$\Delta M^{(em)}$} &
\multicolumn{1}{c}{$\Delta M^{(me)}$} &
\multicolumn{1}{c}{$\Delta M^{(mt)}$} \\[1ex]
\hline
 VId01  &   0.38496~(108) &   0.0002512~(167) &    81.810~(140) &   0.01307~( 27) \\
 VId02  &   0.38884~( 73) &   0.0000244~( 29) &   170.126~( 91) &   0.00169~(  9) \\
 VId03  &  -0.09728~( 10) &  -0.0000264~(  4) &     4.054~(  3) &  -0.00212~(  1) \\
 VId04  &   0.31560~( 20) &   0.0000499~( 20) &    10.720~( 14) &   0.00448~(  5) \\
 VId05  &  -0.02779~( 18) &   0.0000489~( 39) &    16.903~( 20) &   0.00158~(  6) \\
 VId06  &  -0.19938~( 42) &  -0.0001046~( 46) &   -78.953~( 47) &  -0.00548~( 10) \\
 VId07  &  -0.13469~(149) &  -0.0000488~( 66) &   -53.367~( 97) &  -0.00240~( 21) \\
 VId08  &   0.16511~( 99) &   0.0000495~( 14) &   140.628~( 93) &   0.00268~(  8) \\
 VId09  &   0.66926~(342) &   0.0002166~(212) &   160.196~(211) &   0.01340~( 54) \\
 VId10  &  -0.86487~(129) &  -0.0003050~( 90) &  -240.503~(138) &  -0.01865~( 24) \\
 VId11  &   0.19659~(187) &   0.0000178~(131) &    82.026~(122) &   0.00213~( 33) \\
 VId12  &   0.68550~(148) &   0.0003757~(439) &   104.960~(170) &   0.02205~( 61) \\
 VId13  &  -0.84151~(119) &  -0.0002828~( 88) &  -236.280~(142) &  -0.01644~( 22) \\
 VId14  &  -0.55165~(292) &  -0.0000066~(105) &  -241.783~(166) &  -0.00146~( 37) \\
 VId15  &   0.40793~(247) &   0.0002465~(180) &    74.465~(151) &   0.01367~( 43) \\
 VId16  &  -0.86278~(165) &  -0.0002389~(100) &  -194.970~(108) &  -0.01590~( 31) \\
 VId17  &   0.72320~(103) &   0.0000980~( 53) &   335.970~(402) &   0.00704~( 14) \\
 VId18  &  -0.12727~(164) &   0.0002904~( 71) &  -105.046~( 87) &   0.01107~( 22) \\
 VId19  &  -0.68392~(101) &  -0.0002279~(214) &  -142.105~( 98) &  -0.01381~( 36) \\
 VId20  &   0.51681~(295) &   0.0001113~( 72) &   293.527~(151) &   0.00712~( 33) \\
 VId21  &  -0.84441~( 81) &  -0.0003483~( 23) &  -103.944~( 70) &  -0.02040~(  8) \\
 VId22  &   0.32927~(138) &   0.0000666~( 71) &    37.656~( 85) &   0.00574~( 20) \\
 VId23  &   0.30074~(175) &   0.0002004~(126) &    90.028~(135) &   0.01021~( 31) \\
 VId24  &  -0.06021~( 14) &   0.0000016~(  4) &    11.053~( 10) &  -0.00019~(  2) \\
 VId25  &   0.20753~( 13) &   0.0000311~(  3) &    18.408~( 11) &   0.00277~(  1) \\
 VId26  &   0.55652~( 97) &   0.0001401~( 44) &    30.649~( 32) &   0.01020~( 12) \\
 VId27  &   0.04473~( 69) &  -0.0000219~( 31) &    17.953~( 21) &  -0.00182~(  9) \\
 VId28  &   0.19223~(194) &   0.0001718~(146) &    78.298~(170) &   0.00875~( 33) \\
 VId29  &  -0.55884~(161) &  -0.0001263~( 71) &  -306.257~(141) &  -0.00673~( 21) \\
 VId30  &   0.23422~( 35) &   0.0002785~( 65) &    29.833~( 34) &   0.01403~( 11) \\
 VId31  &   0.13852~( 41) &   0.0001240~( 86) &    50.343~( 58) &   0.00596~( 12) \\
 VId32  &  -0.65225~(160) &  -0.0002349~( 53) &  -287.636~(156) &  -0.01276~( 20) \\
 VId33  &   0.30989~(145) &   0.0002517~( 96) &    38.190~(111) &   0.01293~( 21) \\
 VId34  &  -0.74797~(115) &  -0.0002863~( 61) &  -158.487~(118) &  -0.01729~( 17) \\
 VId35  &   0.36986~(130) &   0.0002577~( 77) &   229.607~(120) &   0.01267~( 23) \\
 VId36  &  -0.69348~( 77) &  -0.0002695~( 67) &   -81.830~( 36) &  -0.01390~( 14) \\
 VId37  &  -0.40728~( 48) &  -0.0001806~( 87) &   -80.159~( 48) &  -0.01090~( 17) \\
 VId38  &   0.59921~(159) &   0.0001969~( 44) &   357.390~(157) &   0.01081~( 16) \\
 VId39  &  -0.46380~(126) &  -0.0002631~( 77) &   -78.155~( 82) &  -0.01385~( 21) \\
 VId40  &   0.11874~( 62) &   0.0000125~( 33) &    12.521~( 25) &   0.00079~(  9) \\
 VId41  &  -0.07805~( 16) &  -0.0000071~(  4) &     9.202~( 11) &  -0.00079~(  2) \\
 VId42  &   0.11040~( 83) &  -0.0001267~( 49) &    21.555~( 22) &  -0.00420~( 12) \\
 VId43  &  -0.24138~( 78) &  -0.0000597~( 39) &    18.378~( 39) &  -0.00597~( 11) \\
 VId44  &  -0.12923~(169) &   0.0000835~( 48) &  -123.103~( 97) &   0.00486~( 20) \\
 VId45  &   0.37295~( 35) &   0.0003013~( 64) &    23.573~( 31) &   0.01616~( 11) \\
 [1ex] \hline 
  total &  -0.92943~(941) &   0.0007324~(757) &    37.443~(801) &   0.03080~(158)  
    \end{tabular}
  \end{ruledtabular}
\end{table*}

\endgroup

\begingroup
\squeezetable
\renewcommand{\baselinestretch}{1.2}

\begin{table*}
  \caption{%
Numerical values of diagrams of Set VI(g) 
for the mass-independent contributions $(e,e)$ 
and the mass-dependent contributions $(e,m)$, $(m,e)$, and $(m,t)$. 
The numerals in the parentheses show the uncertainty of the numerical 
integration carried out using VEGAS. 
The number of sampling points is $10^9$ times approximately 100 iterations. 
  \label{table:set6g}
}
  \begin{ruledtabular} \begin{tabular}{ldddd}
\multicolumn{1}{c}{Diagram} & 
\multicolumn{1}{c}{~~$\Delta M^{(ee)}$} &
\multicolumn{1}{c}{~~~$\Delta M^{(em)}$} &
\multicolumn{1}{c}{  $\Delta M^{(me)}$} &
\multicolumn{1}{c}{~~$\Delta M^{(mt)}$} \\[1ex]
\hline
 VIg01  &  -0.63173~( 62) &   0.0008231~( 30) &   -52.384~( 64) &   0.01498~( 11) \\
 VIg02  &   0.31956~( 36) &  -0.0001824~( 14) &    34.460~( 47) &  -0.00257~(  5) \\
 VIg03  &   0.98647~( 72) &   0.0009548~( 12) &   130.839~( 34) &   0.04157~(  6) \\
 VIg04  &  -0.29176~( 64) &  -0.0002605~(  8) &   -81.920~( 72) &  -0.01137~(  5) \\
 VIg05  &  -2.19013~(217) &   0.0001161~(105) &  -185.978~(166) &  -0.02197~( 39) \\
 VIg06  &   0.72494~(122) &   0.0005969~( 46) &   116.641~( 99) &   0.02417~( 17) \\
 VIg07  &  -2.30397~(241) &  -0.0013244~( 98) &  -178.720~(160) &  -0.06282~( 37) \\
 VIg08  &   1.00663~(111) &   0.0004722~( 48) &   109.909~( 85) &   0.01977~( 16) \\
 VIg09  &   0.77315~(138) &   0.0000200~( 54) &   117.911~(124) &   0.00743~( 19) \\
 VIg10  &  -1.00933~(221) &  -0.0004571~(123) &  -182.581~(183) &  -0.01757~( 43) \\
 VIg11  &   1.84018~( 81) &   0.0013280~( 19) &   275.717~( 58) &   0.04634~(  9) \\
 VIg12  &  -1.22515~( 69) &  -0.0003466~( 12) &  -179.818~( 84) &  -0.02208~(  6) \\
 VIg13  &   2.96666~( 99) &   0.0003162~( 20) &   271.391~( 71) &   0.03606~( 10) \\
 VIg14  &  -1.30518~( 79) &   0.0001885~( 14) &  -172.175~( 72) &  -0.00348~(  6) \\
 VIg15  &  -0.60441~( 83) &  -0.0003888~( 13) &  -182.323~( 83) &  -0.01435~(  6) \\
 VIg16  &   1.99802~(100) &  -0.0007980~( 22) &   275.112~( 62) &  -0.01361~( 10) \\
 VIg17  &  -1.85498~(142) &  -0.0017517~( 48) &   -87.674~( 94) &  -0.07228~( 20) \\
 VIg18  &   0.81225~( 82) &  -0.0000606~( 21) &    58.248~( 58) &   0.00235~(  8) \\
 VIg19  &   0.06430~(143) &   0.0015046~( 46) &   -88.590~( 94) &   0.05249~( 20) \\
 VIg20  &   0.11993~( 77) &  -0.0003910~( 17) &    65.647~( 60) &  -0.01272~(  8) \\
 VIg21  &   0.55442~(100) &   0.0004771~( 22) &    58.433~( 83) &   0.01918~( 10) \\
 VIg22  &  -1.87621~(202) &  -0.0002371~( 60) &   -88.324~(126) &  -0.02794~( 25) \\
 VIg23  &   0.72649~( 53) &   0.0009910~( 23) &    57.310~( 59) &   0.03702~(  9) \\
 VIg24  &  -1.42022~(129) &  -0.0018534~( 60) &  -101.191~(102) &  -0.06842~( 19) \\
 VIg25  &  -0.60245~(102) &   0.0002251~( 16) &  -164.454~( 84) &   0.00559~(  9) \\
 VIg26  &   0.75452~(110) &  -0.0007759~( 27) &   258.476~( 81) &  -0.02489~( 13) \\
[1ex] \hline 
  total &   1.66799~(642) &   0.0008141~(286) &     3.961~(489) &   0.06914~( 93)  
    \end{tabular}
  \end{ruledtabular}
\end{table*}

\endgroup

\begingroup
\squeezetable
\renewcommand{\baselinestretch}{1.2}

\begin{table*}
  \caption{%
Numerical values of diagrams of Set VI(h) 
for the mass-independent contributions $(e,e)$ 
and the mass-dependent contributions $(e,m)$, $(m,e)$, and $(m,t)$. 
The numerals in the parentheses show the uncertainty of the numerical 
integration carried out using VEGAS. 
The number of sampling points is $10^9$ times approximately 100 iterations. 
  \label{table:set6h}
}
  \begin{ruledtabular}
    \begin{tabular}{ldddd }
\multicolumn{1}{c}{Diagram} & 
\multicolumn{1}{c}{~~$\Delta M^{(ee)}$} &
\multicolumn{1}{c}{~~~$\Delta M^{(em)}$} &
\multicolumn{1}{c}{  $\Delta M^{(me)}$} &
\multicolumn{1}{c}{~~$\Delta M^{(mt)}$} \\[1ex]
\hline
 VIh01  &   0.35996~( 46) &  -0.0009939~(  6) &   126.906~( 17) &  -0.04544~(  4) \\
 VIh02  &   0.27086~( 37) &   0.0006152~(  4) &   -73.046~( 10) &   0.03092~(  3) \\
 VIh03  &  -0.59951~( 47) &  -0.0007719~(  4) &    50.488~(  9) &  -0.03977~(  3) \\
 VIh04  &   0.26801~( 45) &   0.0001968~(  4) &   -95.157~( 28) &   0.01264~(  3) \\
 VIh05  &   0.13007~( 35) &   0.0000478~(  6) &     1.319~(  2) &   0.00311~(  3) \\
 VIh06  &   0.42117~( 65) &   0.0000669~( 10) &    69.018~( 48) &   0.00532~(  5) \\
 VIh07  &   3.15574~( 62) &   0.0010860~(  7) &   141.662~( 20) &   0.06653~(  5) \\
 VIh08  &   0.40673~( 50) &   0.0000576~( 10) &    41.392~( 15) &   0.00419~(  5) \\
 VIh09  &   0.02536~( 48) &   0.0000732~(  8) &    17.187~( 22) &   0.00327~(  4) \\
 VIh10  &  -2.49371~( 74) &  -0.0010322~( 10) &  -198.433~( 40) &  -0.05948~(  6) \\
 VIh11  &  -0.08174~( 71) &   0.0001225~( 15) &    42.256~( 32) &   0.00482~(  8) \\
 VIh12  &  -1.61239~(108) &  -0.0011801~( 32) &   -94.032~( 71) &  -0.06307~( 16) \\
 VIh13  &   3.12210~( 86) &   0.0000566~( 13) &   262.754~( 32) &   0.01744~( 10) \\
 VIh14  &  -3.55461~(134) &  -0.0011428~( 30) &  -209.006~(113) &  -0.07210~( 14) \\
 VIh15  &   5.66905~( 81) &   0.0021336~( 13) &   273.980~( 33) &   0.12779~( 10) \\
 VIh16  &   0.64030~( 76) &  -0.0002009~( 22) &    78.458~( 33) &  -0.00650~( 11) \\
 VIh17  &   0.46696~( 72) &   0.0001387~( 20) &    35.527~( 46) &   0.00877~( 10) \\
 VIh18  &   0.41308~( 68) &   0.0001996~( 15) &    46.562~( 34) &   0.01193~(  8) \\
 VIh19  &  -2.53838~( 76) &  -0.0008088~( 15) &  -168.962~( 29) &  -0.05015~(  9) \\
 VIh20  &   1.06872~(116) &   0.0008938~( 31) &   -86.585~( 72) &   0.04727~( 16) \\
 VIh21  &   2.24221~( 70) &   0.0009253~( 14) &   117.262~( 29) &   0.05466~(  8) \\
 VIh22  &   0.06791~( 81) &  -0.0001971~( 15) &   143.881~( 80) &  -0.00920~(  9) \\
 VIh23  &  -1.86591~( 44) &  -0.0001763~(  4) &  -212.603~( 17) &  -0.01768~(  3) \\
 VIh24  &  -0.03105~( 50) &   0.0000472~( 12) &    -3.201~(  4) &   0.00194~(  6) \\
 VIh25  &  -0.30141~( 80) &  -0.0001359~( 21) &    51.665~( 47) &  -0.00818~( 12) \\
 VIh26  &  -2.02809~( 77) &  -0.0004662~( 12) &  -198.376~( 59) &  -0.03184~(  8) \\
 VIh27  &   0.92582~( 72) &   0.0006938~( 12) &   -98.570~( 41) &   0.03794~(  8) \\
 [1ex] \hline 
  total &   4.54722~(393) &   0.0002482~( 90) &    62.345~(231) &   0.03516~( 44)  
    \end{tabular}
  \end{ruledtabular}
\end{table*}

\endgroup

\begingroup
\renewcommand{\baselinestretch}{1.2}

\begin{table}
  \caption{%
Auxiliary integrals for Sets VI(d), VI(g), and VI(h). 
The $g\!-\!2$ contribution from the sixth-order vertex diagram
containing a light-by-light scattering subdiagram $a_{6}^{(\ell_1\ell_2)}$ is analytically known for arbitrary combinations of leptons $(\ell_1,\ell_2)$
\cite{Laporta:1991zw,Laporta:1992pa}. 
The numerical values of mass-dependent terms $(e,m)$, $(m,e)$, and $(m,t)$ are
given in \cite{Passera:2006gc}. 
The eighth-order $g\!-\!2$ contributions from Group IV(b) and Group IV(c) 
diagrams 
with the lepton combinations $(e,e)$ and $(m,e)$ ($a_{\text{IVb}}^{(ee)}$, $a_{\text{IVb}}^{(me)}$, $a_{\text{IVc}}^{(ee)}$, and $a_{\text{IVc}}^{(me)}$) are quoted from \cite{Kinoshita:2004wi, Kinoshita:2005zr}.
Other mass-dependent terms of these diagrams, 
$a_{\text{IVb}}^{(em)}$, $a_{\text{IVb}}^{(mt)}$,
$a_{\text{IVc}}^{(em)}$, and $a_{\text{IVc}}^{(mt)}$,
are newly evaluated in this paper.
The forth- and second- order renormalization constants $\Delta\LB_4$ and
$\Delta\LB_2$ are related to our previous  notation through 
$\Delta\LB_4= \Delta L_4 + \Delta B_4$ and $\Delta\LB_2 = \Delta B_2$ 
\cite{Kinoshita:2005zr}.
  \label{table:residual}
}
  \begin{ruledtabular}
  \begin{tabular}{ld@{\extracolsep{6em}}l@{\extracolsep{-4em}}d}
\multicolumn{1}{c}{Integral} & 
\multicolumn{1}{c}{~~~~~~~~~~Value (error)} &
\multicolumn{1}{c}{Integral} & 
\multicolumn{1}{c}{~~~~~~~~~~Value (error)} \\[1ex]
\hline
$a_{6}^{(ee)}$  &   0.371005292\dots  & 
$a_{6}^{(em)}$  &   1.439445989~(77) \times 10^{-5}  \\ 
$a_{6}^{(me)}$  &   20.94792489(16)   & 
$a_{6}^{(mt)}$  &   0.00214283~(69) \\ 
$a_{\text{IVb}}^{(ee)}$ &   0.82249~(28)   & 
$a_{\text{IVb}}^{(em)}$ &   0.00004105~(93) \\
$a_{\text{IVb}}^{(me)}$ &  -0.41704~(375)  & 
$a_{\text{IVb}}^{(mt)}$ &   0.006106~(31)  \\
$a_{\text{IVc}}^{(ee)}$ &  -1.13891~(35)   & 
$a_{\text{IVc}}^{(em)}$ &  -0.0001897~(63) \\
$a_{\text{IVc}}^{(me)}$ &   2.90722~(444)  & 
$a_{\text{IVc}}^{(mt)}$ &  -0.018233~(106) \\
$\Delta\LB_{4}$ &   0.027930~(28)    & 
$\Delta\LB_{2}$ &   0.75           \\
    \end{tabular}
  \end{ruledtabular}
\end{table}

\endgroup


\section{Summary and discussions}
\label{sec:discussion}

In this paper we evaluated the tenth-order QED corrections to 
the anomalous magnetic moments of electron and muon 
from the sets of diagrams, VI(d), VI(g), and VI(h). 

For the electron \gtwo, the total contribution is 
the sum of the mass-independent terms 
(\ref{eq:a1-set6d}), (\ref{eq:a1-set6g}), and~(\ref{eq:a1-set6h}) 
and the mass-dependent terms involving the muon loops 
(\ref{eq:a2-em-set6d}), (\ref{eq:a2-em-set6g}), and~(\ref{eq:a2-em-set6h}):
\begin{align}
  \label{eq:a_e-set6d}
  a_e [\text{Set VI(d)}] &= \ph 1.841~8~(95)  \left(\frac{\alpha}{\pi}\right)^5, \\
  \label{eq:a_e-set6g}
  a_e [\text{Set VI(g)}] &=    -1.591~8~(65)  \left(\frac{\alpha}{\pi}\right)^5, \\ 
  \label{eq:a_e-set6h}
  a_e [\text{Set VI(h)}] &= \ph 0.179~7~(40)  \left(\frac{\alpha}{\pi}\right)^5. 
\end{align}
The tau-lepton contributions
to $a_e$ are more than an order of magnitude smaller than 
Eqs.~(\ref{eq:a2-em-set6d}), (\ref{eq:a2-em-set6g}), and~(\ref{eq:a2-em-set6h})
and lie within the uncertainties of
Eqs.~(\ref{eq:a_e-set6d}), (\ref{eq:a_e-set6g}), and~(\ref{eq:a_e-set6h}).
Thus they are negligible at present.

For the muon  \gtwo, the contributions are the sums of the 
 mass-independent terms 
(\ref{eq:a1-set6d}), (\ref{eq:a1-set6g}), and~(\ref{eq:a1-set6h}) 
and the mass-dependent terms involving electron loops 
(\ref{eq:a2-me-set6d}), (\ref{eq:a2-me-set6g}), and~(\ref{eq:a2-me-set6h})
and tau-lepton loop
(\ref{eq:a2-mt-set6d}), (\ref{eq:a2-mt-set6g}), and~(\ref{eq:a2-mt-set6h}): 
\begin{align}
  \label{eq:a_m-set6d}
  a_\mu [\text{Set VI(d)}] &=    -5.876~(802)  \left(\frac{\alpha}{\pi}\right)^5, \\
  \label{eq:a_m-set6g}
  a_\mu [\text{Set VI(g)}] &= \ph 5.710~(490)  \left(\frac{\alpha}{\pi}\right)^5, \\
  \label{eq:a_m-set6h}
  a_\mu [\text{Set VI(h)}] &=    -8.361~(232)  \left(\frac{\alpha}{\pi}\right)^5. 
\end{align}


\begin{acknowledgments}
This work is supported in part by 
JSPS Grant-in-Aid for Scientific Research (C)19540322 and (C)20540261. 
T. K.'s work is supported in part by the U. S. National Science Foundation
under Grant PHY-0757868,
and the International Exchange Support Grants (FY2010) of RIKEN.
T. K. thanks RIKEN for the hospitality extended to him
while a part of this work is carried out.
The numerical calculation was conducted on 
the RIKEN Super Combined Cluster (RSCC) and 
the RIKEN Integrated Cluster of Clusters (RICC) supercomputing systems.
\end{acknowledgments}

\appendix*

\section{Algorithm for identifying independent loops}
\label{sec:loopmatrix}

This Appendix describes an algorithm for identifying the fundamental set 
of circuits of a diagram following Ref.~\cite{Nakanishi:1971}
which is adapted to automated handling by computers. 

In the parametric integral approach, 
the integrand of the amplitude is expressed in terms of ``building blocks'', 
$B_{ij}$, $U$, $V$, $A_j$, which are functions of Feynman parameters. 
$B_{ij}$ reflects loop structure of the diagram, 
while $A_j$ is related to the flow of the external momenta. 

The definition of $B_{ij}$ is given as follows \cite{Kinoshita:1990}. 
A chain diagram $\widetilde{\mathcal{G}}$ is derived 
from the diagram $\mathcal{G}$ by 
removing all the external lines and 
disregarding the distinction of the type of lines. 
Suppose that a fundamental set of circuits 
(independent self-nonintersecting loops) of the chain diagram is known. 
Then, for $i$ and $j$ that label the lines of $\widetilde{\mathcal{G}}$, 
$B_{ij}$ is given by 
\begin{align}
  \label{eq:def-u}
  U_{st} &= \sum_{k}\,z_k \xi_{k,s} \xi_{k,t}, 
  \quad
  U = \det_{st}\,U_{st}, \\[1ex]
  \label{eq:def-bij}
  B_{ij} &= U \sum_{s,t}\,\xi_{i,s} \xi_{j,t} (U^{-1})_{st}. 
\end{align}
where $s$ and $t$ refer to the circuits. The loop matrix $\xi_{k,c}$ 
takes $(1,-1,0)$ according to whether the line $i$ is (along, against, 
outside of) circuit $c$. 
All lines are assumed to be appropriately directed. 

\begin{figure}
  \includegraphics[scale=1.0]{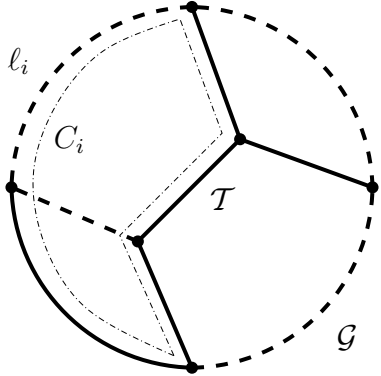}
  \caption{%
    \label{fig:circuit}
    A choice of maximal tree $\mathcal{T}$ on graph $\mathcal{G}$ 
    is shown in solid lines. 
    The chord set $\mathcal{T}^\ast$ consists of the dashed lines. 
    A circuit $C_i$ is found for a line $\ell_i \in \mathcal{T}^\ast$ 
    and a path on $\mathcal{T}$ shown in thick lines 
    that connects both ends of $\ell_i$. 
}
\end{figure}

The circuits are found in the following way. 
A maximal tree $\mathcal{T}$ of a graph $\mathcal{G}$ is a 
simply-connected set of lines that connects all vertices of 
$\mathcal{G}$ and does not have loops. 
An example is shown in Fig.~\ref{fig:circuit} 
where solid lines denote the maximal tree $\mathcal{T}$. 
For any pair of vertices there is a unique path 
on $\mathcal{T}$ that links these vertices, 
because $\mathcal{T}$ is simply-connected and 
if there were more than one path $\mathcal{T}$ would have a loop. 
$\mathcal{T}$ consists of $n_v-1$ lines where $n_v$ is the number of 
vertices of $\mathcal{G}$. 

The chord set (or cotree) $\mathcal{T}^\ast$ is the complement of 
$\mathcal{T}$ against $\mathcal{G}$, i.e., 
$\mathcal{T}^\ast \cap \mathcal{T} = \emptyset$ and 
$\mathcal{T}^\ast \cup \mathcal{T} = \mathcal{G}$. 
In Fig.~\ref{fig:circuit} the lines in $\mathcal{T}^\ast$ are 
shown by dashed lines. 
The number of lines of $\mathcal{T}^\ast$ is $n_p - n_v + 1 = n_l$ 
where $n_p$ and $n_l$ are the number of lines and loops of $\mathcal{G}$, 
respectively. 
Then, for each $\ell_i \in \mathcal{T}^\ast$, 
there is a path $\mathcal{P}_i$ on the tree $\mathcal{T}$ which links two 
end-points of $\ell_i$ as shown above. $\ell_i$ and $\mathcal{P}_i$ 
form a closed loop $C_i$. 
These closed loops are independent with each other by construction, 
namely, $\ell_i \notin C_j$ if $i \ne j$. 
They form a fundamental set of circuits. 

For the diagrams without closed lepton loops (\textit{q-type} diagrams) 
as described in Ref.~\cite{Aoyama:2005kf}, 
the fundamental set of circuits are found rather trivially. 
In this case the maximal tree $\mathcal{T}$ is chosen as the set of 
lines that forms the open lepton line, 
and thus the chord set $\mathcal{T}^\ast$ consists of the photon lines. 
Therefore, each circuit is identified by a photon line and a string 
of lepton lines that connects both ends of the photon line. 

For a general diagram, a maximal tree of the diagram can be found 
in the following way. 
Choose a vertex $v_0$ of $\mathcal{G}$, and starting from $v_0$, 
extend the tree by adding a line adjacent to the vertex of the tree, 
of which the other vertex is not yet included, 
until all vertices of $\mathcal{G}$ are visited. 
During this process a unique path $\mathcal{P}(v_0, v_a)$, i.e.~a string 
of lines in $\mathcal{T}$ that runs from $v_0$ to $v_a$, is also 
found for each vertex $v_a$. 
A path on $\mathcal{T}$ connecting an arbitrary pair of vertices 
$v_a$ and $v_b$ is then found by joining the paths $\mathcal{P}(v_b, v_0)$ 
and $\mathcal{P}(v_0, v_a)$ where 
$\mathcal{P}(v_b, v_0) \equiv \mathcal{P}^{-1}(v_0, v_b)$ 
is the string of lines of $\mathcal{P}(v_0, v_b)$ in reversed order. 
Note that the duplicated lines in the paths are to be eliminated. 


\bibliographystyle{apsrev}
\bibliography{b}

\end{document}